\documentclass[11pt]{article}
\newlength{\vshift}
\newlength{\hshift}
\usepackage{amssymb,amsopn}

\def\ds{\stackrel{\star}{,}}

\def\bea{\begin{eqnarray}}
\def\eea{\end{eqnarray}}
\def\bb{\begin{equation}}
\def\eb{\end{equation}}
\def\Tr{{\rm Tr \,}}
\newcommand{\initiate}{\setcounter{equation}{0}}

\begin{document}

\newpage
\renewcommand{\theequation}{\arabic{section}.\arabic{equation}}
\begin{titlepage}

\begin{center}

{\Large{\bf   Deformed Gauge Theories}}

\vskip 2em

{{\bf Julius Wess${}^{1,2,3}$}}

\vskip 2em
Lecture given at Workshop "Noncommutative Geometry in Field and String Theories", Corfu Summer Institute on EPP, September 2005,  Corfu, Greece.

\vskip 3em

${}^{1}$Arnold Sommerfeld Center for Theoretical Physics\\
Universit\"at M\"unchen, Fakult\"at f\"ur Physik\\
Theresienstr.\ 37, 80333 M\"unchen, Germany\\[1em]

${}^{2}$Max-Planck-Institut f\"ur Physik\\
        F\"ohringer Ring 6, 80805 M\"unchen, Germany\\[1em]

${}^{3}$Universit\"at Hamburg, II Institut f\"ur Theoretische Physik\\
and DESY, Hamburg\\
        Luruper Chaussee 149, 22761 Hamburg, Germany\\[1em]

\end{center}

\vspace{1.5em}

\abstract
{Gauge theories are studied on a space of functions with the Moyal-Weyl product. The development of these ideas follows the differential geometry of the usual gauge theories, but several changes are forced upon us. The Leibniz rule has to be changed such that the theory is now based on a twisted Hopf algebra. Nevertheless, this twisted symmetry structure leads to conservation laws. The symmetry has to be extended from Lie algebra valued to enveloping algebra valued and new vector potentials have to be introduced. As usual, field equations are subjected to consistency conditions that restrict the possible models. Some examples are studied.}

\vspace{2em}
\noindent This article is based on common work with Paolo Aschieri, Christian Blohmann, Marija Dimitrijevi\' c, Branislav Jur\v co, Frank Meyer, Stefan Schraml, Peter Schupp and Michael Wohlgenannt.

\vspace{2em}
{\bf Keywords:} deformed spaces, Hopf algebras, deformed symmetry, noncommutative gauge theory

{\bf PACS}: 02.40.Gh, 02.20.Uw

{\bf MSC}: 81T75 Noncommutative geometry methods, 58B22 Geometry of Quantum groups

\vspace{1.5em}
{E-mail: wess@theorie.physik.uni-muenchen.de}
\vfill

\end{titlepage}

\section{Introduction}

Gauge theories have been formulated and developed on the algebra of 
functions with a pointwise product:
\bb
\mu\{f\otimes g \}= f\cdot g . 
\label{I.1}
\eb
This product is associative and commutative.

Recently, algebras of functions with a deformed product have been
studied intensively \cite{star}. These deformed (star-)products remain associative but
not commutative.

The simplest example is the Moyal-Weyl product.
\bb
\mu_\star\{f\otimes g \}= \mu\{ e^{\frac i2 \theta^{\rho\sigma}\partial_\rho\otimes\partial_\sigma}f\otimes g\} .
\label{I.2}
\eb
It had its first appearance in quantum mechanics \cite{wm}.

The star product can be seen as a higher order $f$-dependent
differential operator acting on the function $g$. For the 
example of the Moyal-Weyl product this is 
\bb
 f\star g = \sum _{n=0}^\infty \frac{1}{n!} \Big(\frac{i}{2}\Big)^n
\theta^{\rho_1\sigma_1}\dots
\theta^{\rho_n\sigma_n}\Big(\partial_{\rho_1}\dots\partial_{\rho_n}f\Big)
 \partial_{\sigma_1}\dots\partial_{\sigma_n} \cdot g  .
\label{I.3}
\eb
The differential operator maps the function $g$ to the 
function $f\star g$.

The inverse map also exists \cite{vienna}, \cite{grav1}. It $\star$-maps the function $g$
to the function obtained by pointwise multiplying it with $f$
\begin{equation}
X^\star_f \star g = f\cdot g
 \label{I.4}
\end{equation}
For the Moyal-Weyl product we obtain
\begin{equation}
X^\star_f = \sum _{n=0}^\infty \frac{1}{n!} \Big(-\frac{i}{2}\Big)^n
\theta^{\rho_1 \sigma_1}\dots
\theta^{\rho_n \sigma_n}\Big(\partial_{\rho_1}\dots\partial_{\rho_n}f\Big)\star
\partial^\star_{\sigma_1}\dots\partial^\star_{\sigma_n}  .
\label{I.5}
\end{equation}

The star-acting derivatives we denote by $\partial^\star_\rho$. 
For the Moyal-Weyl product the $\star$-derivatives and the usual derivatives are the same. In general this will not be the case.
Star differentiation and star differential operators have been thoroughly discussed  \cite{grav1}, \cite{mex}.

In this lecture we are going to study gauge transformations on
Moyal-Weyl or $\theta$-{\it deformed} spaces.

\section{ Gauge transformations}

{\it Undeformed} gauge transformations are Lie algebra valued:
\bea
&& \delta_\alpha \phi (x)= i\alpha(x)\phi(x) , \nonumber\\
&& \alpha (x) =\sum_l \alpha^l(x) T^l  ,\label{II.1}\\
&&  [T^l, T^k] = if^{lkr}T^r  ,\nonumber\\
&&[\delta_\alpha,\delta_\beta ]\phi = [\alpha ,\beta ]\phi = -i\delta_{[\alpha,\beta]}\phi . \nonumber
\eea

The {\it deformed} gauge transformations \cite{gaugeth1}, \cite{gaugeth2} are defined as follows:
\begin{equation}
\delta^\star_\alpha \phi = iX^\star_{\alpha}\star \phi = iX^\star_{\alpha^l} T^l\star \phi
= i\alpha\cdot\psi .
\label{II.2}
\end{equation}
From the fact that $X^\star _f\star X^\star_g = X^\star_{f\cdot g}$ we conclude:
\begin{equation}
[ \delta^\star_\alpha \ds \delta^\star_\beta ] \phi 
 = - i\delta^\star_{[\alpha ,\beta] }\phi .
\label{II.3}
\end{equation}
The $\star$-transformations represent the algebra via the $\star$-commutator.

\vskip0.5cm
Before we construct gauge theories we have to learn how products
of fields transform.

In the {\it undeformed} situation we use, without even thinking, 
the Leibniz rule:
\bb
\delta_\alpha (\phi\cdot \psi)  = (\delta_\alpha\phi )\cdot\psi + \phi\cdot (\delta_\alpha\psi ) 
\label{II.4}
\eb
and we can easily verify that this Leibniz rule is consistent with the 
Lie algebra:
\bb
[\delta_\alpha,\delta_\beta ](\phi\cdot\psi ) = -i\delta_{[\alpha,\beta]}(\phi\cdot\psi ) .
\label{II.5}
\eb

For the {\it deformed} transformation law of a $\star$-product of fields
we demand a transformation law that is in the class of 
transformations defined in (\ref{II.2})  \cite{vienna}, \cite{grav1}, \cite{gaugeth1}, \cite{vbanja}, \cite{chaich}.
This amounts to first decomposing the representation $\phi\star\psi$ for
$x$-independent parameters into its irreducible parts and then 
follow (\ref{II.2}) for gauging
\bb
\delta_\alpha^\star (\phi\star\psi) = i X^\star_{\alpha^l}\star\{ T^l\phi\star\psi+\phi\star T^l\psi\} .
\label{II.6}
\eb
Certainly it is consistent with the Lie algebra:
\bb
[\delta^\star_\alpha \ds\delta^*_\beta ](\phi\star\psi ) =-i\delta^\star_{[\alpha,\beta]}(\phi\star\psi )
\label{II.7}
\eb
Because $\phi\star\psi$ is a function we can use the definition of $X^\star_f$
given in (\ref{I.4}) and simplify (\ref{II.6})
\bb
\delta_\alpha^\star (\phi\star\psi) = i \alpha^l\cdot \{ T^l\phi\star\psi+\phi\star T^l\psi\} .
\label{II.8}
\eb
As $\alpha^l$ does not commute with the $\star$-operation this is different
from (\ref{II.4}).
To see this difference more clearly we expand (\ref{II.8}) in $\theta$
\begin{eqnarray}
\delta^\star_\alpha (\phi\star\psi )  \hspace*{-2mm}&=&\hspace*{-2mm} i\alpha^l \{T^l\phi\cdot \psi + \phi \cdot T^l\psi \nonumber \\
 &+&\hspace*{-2mm}\frac i2 \theta^{\rho\sigma} \Big( T^l\partial_\rho \phi\cdot  \partial_\sigma\psi  + \partial _\rho\phi \cdot T^l\partial_\sigma \psi \Big) +  O(\theta^2)\} .
 \label{II.9}
\end{eqnarray}

The final version of the Leibniz rule for the $\star$-product 
should be entirely expressed with $\star$-operations.
Thus we express (\ref{II.9}) with $\star$-products. A short 
calculation shows:
\begin{eqnarray}
\delta^\star_\alpha (\phi\star\psi )  \hspace*{-2mm}&=&\hspace*{-2mm} i(\alpha \phi)\star\psi + i\phi \star (\alpha \psi) \label{II.10} \\
 &-&\hspace*{-2mm}\frac i2 \theta^{\rho\sigma} \Bigg( 
i\Big( (\partial_\rho \alpha^l)T^l\phi\Big) \star (\partial_\sigma\psi)  + (\partial _\rho\phi)\star i\Big((\partial_\sigma \alpha^l) T^l\psi\Big) \Bigg) +  O(\theta^2) .
\nonumber
\end{eqnarray}
With more work we can prove by induction  to all orders in $\theta$  
 the following equation:
\begin{eqnarray}
\delta^\star_\alpha (\phi\star\psi )  \hspace*{-2mm}&=&\hspace*{-2mm}
i\sum_{n=0}^\infty\frac{1}{n!} \Big( -\frac i2\Big) ^n \theta^{\rho_1\sigma_1}\dots \theta^{\rho_n\sigma_n}
\{(\partial_{\rho_1}\dots\partial_{\rho_n}\alpha)\phi\star (\partial_{\sigma_1}\dots\partial_{\sigma_n}\psi) 
\nonumber\\ &+&
(\partial_{\rho_1}\dots\partial_{\rho_n}\phi)\star(\partial_{\sigma_1}\dots\partial_{\sigma_n}\alpha)\psi \}  .
\label{II.11}
\end{eqnarray}

This is different from what we obtain by putting just stars in 
the Leibniz rule (\ref{II.4}). But this difference has a well-defined
meaning if we use the Hopf algebra language to derive the 
Leibniz rule.

\initiate
\section{Hopf algebra techniques}

The essential ingredient for a Hopf algebra \cite{Hopf} is the comultiplication
$\Delta(\alpha)$: For the {\it undeformed} situation we define:
\bb
\Delta(\alpha ) = \alpha\otimes 1 +1\otimes\alpha .
\label{III.1}
\eb
It allows us to write the Leibniz rule (\ref{II.4}) in the Hopf
algebra language:
\begin{equation}
\delta_\alpha (\phi\cdot\psi) = \mu \{ \Delta(\alpha)\phi\otimes\psi \}  .
 \label{III.2}
\end{equation}
In the {\it deformed} situation we use a twisted coproduct:
\bea
\Delta_{\cal F}(\alpha)  \hspace*{-2mm}&=&\hspace*{-2mm}
{\cal F}(\alpha\otimes 1+ 1\otimes\alpha ){\cal F}^{-1} ,\nonumber \\
{\cal F}  \hspace*{-2mm}&=&\hspace*{-2mm} e^{-\frac i2 \theta^{\rho\sigma}\partial_\rho\otimes\partial_\sigma} .
\label{III.3}
\eea
Here ${\cal F}$ is a twist that has all the properties to define a Hopf algebra with
$\Delta_{\cal F}(\alpha)$ as a comultiplication \cite{twist1}, \cite{twist2}, \cite{twist3}. We can show that 
the transformation (\ref{II.11}) can be written  in the form
\bb
\delta^\star_\alpha (\phi\star\psi)  = i
\mu_\star\{ \Delta_{\cal F}(\alpha )\phi\otimes \psi\} 
 \label{III.4}
\eb
that defines the Leibniz rule in terms of the twisted comultiplication and
the product $\mu_\star$. To show this we start from equation (\ref{II.8}) and write it with the explicit definition of the $\star$-product:
\bea
 \delta_\alpha^\star (\phi\star\psi )\hspace*{-2mm}&=&\hspace*{-2mm} i\alpha^l\mu\{ e^{\frac i2 \theta^{\rho\sigma} \partial_\rho\otimes\partial_\sigma  } (T^l\phi\otimes \psi +\phi\otimes T^l\psi) \}
 \nonumber \\
 \hspace*{-2mm}&=&\hspace*{-2mm}   i\sum_{n=0}^\infty\frac{1}{n!} \Big( \frac i2\Big) ^n \theta^{\rho_1\sigma_1}\dots \theta^{\rho_n\sigma_n}
\Big( \alpha^l T^l (\partial_{\rho_1}\dots\partial_{\rho_n}\phi) (\partial_{\sigma_1}\dots\partial_{\sigma_n}\psi) 
\nonumber\\ 
&& + (\partial_{\rho_1}\dots\partial_{\rho_n}\phi)\alpha^l T^l(\partial_{\sigma_1}\dots\partial_{\sigma_n}\psi) \Big)  .
\label{III.4'}
\eea
This we now rewrite as follows
\bea
 \delta_\alpha^\star (\phi\star\psi )\hspace*{-2mm}&=&\hspace*{-2mm} i\mu
(\alpha\otimes 1 + 1\otimes\alpha)
 e^{\frac i2 \theta^{\rho\sigma} \partial_\rho\otimes\partial_\sigma  }\phi\otimes \psi 
 \nonumber \\
 \hspace*{-2mm}&=&\hspace*{-2mm}   i\mu\{  e^{\frac i2 \theta^{\rho\sigma} \partial_\rho\otimes\partial_\sigma  } \cdot
 e^{-\frac i2 \theta^{\rho\sigma} \partial_\rho\otimes\partial_\sigma  }(\alpha\otimes 1 + 1\otimes\alpha)
 e^{\frac i2 \theta^{\rho\sigma} \partial_\rho\otimes\partial_\sigma  } \phi\otimes\psi \nonumber \\
\hspace*{-2mm}&=&\hspace*{-2mm}  i\mu_\star\{ \Delta_{\cal F} (\alpha)\phi\otimes\psi \} .\label{III.4''}
\eea

\vskip 0.5cm

The Hopf algebra of gauge transformations can also be formulated with 
functional derivatives.
We again start with the gauge transformation in the {\it undeformed}
situation.
\bea
\delta_\alpha\phi_r(x)  \hspace*{-2mm}&=&\hspace*{-2mm} i\alpha_{rk}(x)\phi_k(x)
\label{III.5}  \\
 \hspace*{-2mm}&=&\hspace*{-2mm} i\int dz\alpha_{jk}(z)\phi_k(z)\frac{\delta}{\delta\phi_j(z)}\cdot\phi_r(x) . \nonumber
\eea
The fields $\phi$ can be in a reducible representation as well.

The generators of gauge transformations
\bb
S_\alpha =i\int dz \alpha_{jk}(z)\phi_k(z)\frac{\delta}{\delta\phi_j(z)}
\label{III.6}
\eb
can be considered as vector fields in the space of fields. They
represent the algebra:
\bb
[S_\alpha,S_\beta ] =-i S_{[\alpha,\beta ]} .
\label{III.7}
\eb

A Hopf algebra structure can now be introduced via the coproduct:
\bb
\Delta(S_\alpha) = S_\alpha\otimes 1 + 1\otimes S_\alpha .
\label{III.8}
\eb
It is easy to verify that it is consistent with the algebra:
\bb
[\Delta (S_\alpha),\Delta (S_\beta) ] =-i \Delta (S_{[\alpha,\beta ]})
\label{III.9}
\eb
and leads to the Leibniz rule
\bb
\mu\{\Delta (S_\alpha )\phi\cdot\psi \} = i\Big( (\alpha\phi )\cdot\psi +\phi\cdot (\alpha\psi )\Big) .
\label{III.10}
\eb

Again, we can deform the coproduct by a twist:
\bb
{\cal F} = e^{-\frac i2 \theta^{\mu\nu}\int dz\partial_\mu\phi_l(z)\cdot\frac{\delta}{\delta\phi_l(z)}\otimes
\int dy\partial_\nu\phi_k(y)\cdot\frac{\delta}{\delta\phi_k(y)}}
\label{III.11}
\eb
and define
\bb
\Delta_{\cal F}(S_\alpha) ={\cal F}^{-1}\Delta (S_\alpha){\cal F} .
\label{III.12}
\eb

This twisted coproduct is again compatible with the Hopf algebra
structure. When we derive the Leibniz rule from it
\bb
\delta^\star_\alpha (\phi_r\star\phi_s ) =\mu_\star\{\Delta_{\cal F}(S_\alpha)\phi_r\otimes\phi_s \}
\label{III.13}
\eb
we obtain (\ref{II.11}). The Leibniz rules are identical. 

The advantage of this formulation is that it is easy to include gauge fields 
as well.  In the {\it undeformed} situation they are Lie algebra valued and transform as follows:
\bb
\delta A_\mu = \partial_\mu\alpha + i\alpha^l[T^l,A_\mu ] .
\label{III.14}
\eb
This gives rise to an additional term in the generator $S_\alpha$:
\bb
S_\alpha^{A_\mu} =\int dz\{\partial_\mu\alpha^l(z)-\alpha^r(z)A^s_\mu(z)f^{rsl}\} \frac{\delta}{\delta A_\mu^l(z)} .
\label{III.15}
\eb
It generates the gauge transformations of $A_\mu^l(z)$ and it is
consistent with the algebra relation (\ref{III.7}). In the
coproduct it has to be included and
for the twist it demands an additional term as well:
\bb
{\cal F} = e^{-\frac i2 \theta^{\mu\nu}\int dz\Big( \partial_\mu\phi^l\frac{\delta}{\delta\phi_l}+\partial_\mu A_\rho^l \frac{\delta}{\delta A_\rho^l} \Big)\otimes
\int dy\Big( \partial_\nu\phi^l\frac{\delta}{\delta\phi_l}+\partial_\nu A_\rho ^l\frac{\delta}{\delta A_\rho^l} \Big)} .
\label{III.16}
\eb

We can now calculate the contribution of the gauge field to the
Leibniz rule. As an example we calculate:
\bb
\delta_\alpha^\star (A_\mu\star\phi ) =\mu^\star \{ \Delta_{\cal F}(\alpha )A_\mu\otimes \phi \}
\label{III.17}
\eb
and obtain:
\bea
\delta^\star_\alpha(A_\mu\star\psi ) \hspace*{-2mm}&=&\hspace*{-2mm}i\alpha^l\Big( [T^l,A_\mu ]\star\psi\Big) 
+  i\alpha^l\Big( A_\mu\star T^l\psi\Big) + (\partial_\mu)\alpha^l T^l\psi
 \nonumber \\
\hspace*{-2mm}&=&\hspace*{-2mm}i\alpha^l T^l(A_\mu\star\psi ) + (\partial_\mu)\alpha\psi .
 \label{III.18}
\eea

Now we can define a covariant derivative
\bb
D^\star_\mu\psi = \partial_\mu\psi -iA_\mu\star\psi .
\label{III.19}
\eb
It will transform as usual if  the vector field $A_\mu$  transforms as in (\ref{III.14}):
\bea
 &&\delta^\star_\alpha (D_\mu^\star\psi )= i\alpha^lT^l(D_\mu^\star\psi )=
 iX^\star_{\alpha^l}\star T^l(D^\star_\mu\psi ) ,
\label{III.20}  \\
&& \delta^\star_\alpha A_\mu =  \partial_\mu \alpha +i\alpha^l[T^l,A_\mu]
=\partial_\mu\alpha +iX^\star_{\alpha^l}\star [ T^l,A_\mu] .
\label{III.21}
\eea

From (\ref{III.21}) we see that a Lie algebra valued
vector field remains Lie algebra valued by the transformation (\ref{III.21}).

\section{Field equations}

Now we proceed as in the {\it undeformed} situation. First we
define the field strength tensor:
\bea
F_{\mu\nu}
\hspace*{-2mm}&=&\hspace*{-2mm} i[D^\star_\mu\ds D^\star_\nu]  \label{IV.1}
\\
\hspace*{-2mm}&=&\hspace*{-2mm}\partial_\mu A_\nu -\partial_\nu A_\mu -i[A_\mu\ds A_\nu ]  .\nonumber
\eea
Here we see already that $F_{\mu\nu}$ will not be Lie algebra valued even for
Lie algebra valued vecotr fields. 

Using the twisted gauge transformations
we derive the transformation law of the fieldstrangth tensor: 
\bb
\delta_\alpha^\star F_{\mu\nu} = iX^\star_{\alpha^l}\star [T^l,F_{\mu\nu}] = i[\alpha, F_{\mu\nu}] .
\label{IV.2}
\eb
The expression $F^{\mu\nu}\star F_{\mu\nu}$ will transform accordingly.

The invariant Lagrangian we define as usual:
\bb
{\cal L} = \frac 1c \, \Tr (F^{\mu\nu}\star F_{\mu\nu}) .
\label{IV.3}
\eb
It is invariant and it is a deformation of the {\it undeformed} Lagrangian
of a gauge theory.

To speak about an action we have to define integration. We take the 
usual integral over $x$ and can verify that 
\begin{equation}
\int d^4x\,  f\star g  = \int d^4 x\,  g\star f   =\int d^4 x\,  f\cdot g 
\label{IV.4}
\end{equation}
by partial integration. This is called the trace property of the integral.

Equation (\ref{IV.4}) allows a cyclic permutation of the fields under the integral.
To derive the field equations we take the field to be varied to the very 
left. We work with the action
\bb
S = \frac 1c \int  d^4x \,\Tr (F^{\mu\nu}\star F_{\mu\nu}) .
\label{IV.5}
\eb
From the trace property we compute:
\bea
\frac{\delta S}{\delta A_\rho (z)}
\hspace*{-2mm}&=&\hspace*{-2mm}\frac 2c \int d^4x\,\Tr \frac{\delta F_{\mu\nu}(x)}{\delta A_\rho(z)} \star F^{\mu\nu}(x)
\label{IV.6} \\
\hspace*{-2mm}&=&\hspace*{-2mm}
\frac 2c \int d^4x\, \Tr \frac{\delta }{\delta A_\rho(z)} (\partial_\mu A_\nu -\partial_\nu A_\mu -i[A_\mu\ds A_\nu ] )\star F^{\mu\nu}(x)  \nonumber \\
\hspace*{-2mm}&=&\hspace*{-2mm}\frac 4c \int d^4x\,\Tr \frac{\delta }{\delta A_\rho(z)} (\partial_\mu A_\nu -iA_\mu\star A_\nu  )\star F^{\mu\nu}(x)  \nonumber
\eea
because $F^{\mu\nu}$ is antisymmetric.
The last term is obtained after a cyclic permutation. The field equations
are
\bb
\frac{\delta S}{\delta A_\rho (z)} =\frac 4c \int d^4 x\Tr \frac{\delta A_\nu(x) }{\delta A_\rho(z)}\star (-\partial _\mu F^{\mu\nu}(x) 
+iA_\mu\star F^{\mu\nu} - i F^{\mu\nu}\star A_\mu ) .
\label{IV.7}
\eb
These are exactly the equations we have expected from covariance:
\bb
D_\mu^\star F^{\mu\nu} = \partial_\mu F^{\mu\nu} - i[A_\mu \ds F^{\mu\nu}] = 0 .
\label{IV.8}
\eb

We have already seen that $F_{\mu\nu}$ cannot be Lie algebra valued.
From the field equations, considered as equations for the 
vector potential $A_\mu$ we see that $A_\mu$ cannot be Lie algebra
valued either. We have to consider $F_{\mu\nu}$ and $A_\mu$  to be enveloping algebra valued. The additional vector field will introduce additional ghosts in the Lagrangian. To eliminate them we have to enlarge the symmetry to be enveloping algebra valued as well. For simplicity we assume  $\alpha$, $A_\mu$ and $F_{\mu\nu}$ to be matrix valued when the matrices act in the representation space of $T^l$.

From the field equations (\ref{IV.8}) follows a consistency equation because
$F^{\mu\nu}$ is antisymmetric in $\mu$ and $\nu$:
\bb
\partial_\nu [A_\mu\ds F^{\mu\nu}] =0 .
\label{IV.99}
\eb
To verify this condition we have to use the field equations:
\bb
\partial_\nu [A_\mu\ds F^{\mu\nu}] = [\partial_\nu A_\mu \ds F^{\mu\nu}]+[A_\mu\ds \partial_\nu F^{\mu\nu}]
\label{IV,10}
\eb
In the first term we replace $\partial_\nu A_\mu$ by $ \frac 12(\partial_\nu A_\mu -\partial_\mu A_\nu )$ because $F_{\mu\nu}$ is antisymmetric in
$\mu$ and $\nu$. Then we express this term by $F_{\mu\nu}$ according to ({\ref{IV.1}):
\bb
\frac 12 (\partial_\nu A_\mu -\partial_\mu A_\nu)=\frac i2 F_{\nu\mu} + \frac i2 [A_\nu\ds A_\mu ] .
\label{IV,11}
\eb

The $[F^{\mu\nu}\star F_{\mu\nu}]$ commutator vanishes and we are left with
 $ \frac i2[[A_\nu\ds A_\mu]\ds F^{\mu\nu}]$ for the 
first term in (\ref{IV,10}).
For the second term in (\ref{IV,10}) we use the field equations (\ref{IV.8}).
Finally all terms left add up to zero if we use the Jacobi identity.

In all these equations $A_\mu$ and $F_{\mu\nu}$ are supposed to be matrices. We have
supressed the matrix indices.
 
A conserved current was found
\bb
j^\nu =[A_\mu\ds F^{\mu\nu}], \qquad \partial_\nu j^\nu = 0 .
\label{IV,12}
\eb
For $\theta =0$ this is the current of a non-abelian gauge theory.

\initiate
\section{ Matter fields}

Matter fields can be coupled covariantly to the gauge fields via
a covariant derivative. We start from a multiplet of the gauge 
group $\psi_A$ not necessarily irreducible. The index $A$ denotes the
component of the field $\psi$ in the representation space. The transformation
law of $\psi$ is:
$\delta\psi_A =iX^\star_{\alpha_{AB}^\star}\psi_B = i\alpha_{AB}\psi_B$.
For the usual gauge transformations $\alpha_{AB}$ will be Lie algebra valued. The covariant 
derivative is:
\bb
(D_\mu^\star \star \psi)_A = \partial_\mu\psi_A -i A_{\mu AB}\star \psi_B .
\label{V,1}
\eb
The gauge potential $A_\mu$ in now supposed to be matrix valued in the representation space 
spanned by the matter fields. 

For a spinor field
\bb
\bar\psi_{\alpha A} \gamma^\mu_{\alpha\beta}(D^\star_\mu \star \psi)_A
\label{V,2}
\eb
will be invariant and therefore suitable for a covariant Lagrangian.

We consider the Lagrangian:
\bb
{\cal L} = \frac 1c \Tr F^{\mu\nu}F_{\mu\nu} + \bar\psi\star \gamma^\mu (i\partial_\mu  + A_\mu \star )\psi -m\bar{\psi}\star\psi .
\label{V,3}
\eb
We have suppressed the matrix indices.

The field equations are obtained from (\ref{V,3}) by varying the fields:
\bb
\frac{\delta{\cal L}}{\delta A_\rho}=\partial_\mu F^{\mu\rho}_{AB} +i[A_{\mu }\ds F^{\rho\mu}_{ }]_{AB} 
+\gamma^\rho_{\alpha\beta}\psi_{\beta A}\star \bar\psi_{\alpha B} = 0 \label{V,4}
\eb
and for the matter fields:
\bea
\frac{\delta{\cal L}}{\delta\bar\psi} \hspace*{-2mm}&=&\hspace*{-2mm} \gamma^\mu(\partial_\mu\psi_A -iA_{\mu AB}\star\psi_B ) + im\psi_A = 0 \label{V,5}\\
\frac{\delta{\cal L}}{\delta\psi} \hspace*{-2mm}&=&\hspace*{-2mm}(\partial_\mu\bar\psi_A \gamma^\mu 
+ i\bar\psi_B\gamma^\mu\star iA_{\mu AB} ) - im\bar\psi_A = 0 .
\nonumber
\eea

Again, equation (\ref{V,4}) leads to a consistency relation that can be
verified with the help of the field equations. It is, however,
important that the representation space for the field $\psi$ and the 
vector potential $A_{\mu AB}$ are the same. The representation space of the matter fields determines the space for the gauge potentials.

We conclude that   there is a conserved
current:
\bb
j^\rho_{AB} = i[A_\mu\ds F^{\mu\rho}]_{AB} -\gamma^\rho_{\alpha\beta}\psi_{\beta A}\star\bar\psi_{\alpha B} .
\label{V,6}
\eb

We were again able to find a conserved current as a consequence of a deformed
symmetry. Even if we put the vector potential to zero there remains
the part from the matter field. There are conservation laws due to a deformed 
symmetry. It is remarkable that we have found conserved currents in the twisted theory as well. In the {\it undeformed} theory we can derive them with the help of the Noether theorem. In the deformed theory this is not possible. Nevertheless the property that a theory has a conserved current is preserved by a deformation. This is an important step to convince ourselves that a deformed gauge theory has properties close to what we need for physics.

\section{ Examples}

{\bf 1) Maxwell equations}

We start from the simplest gauge theory based on U(1) and describing
gauge fields only. We proceed schematically:
The transformation law of the gauge field $A_\mu$:
\bb
\delta_\alpha A_\mu =\partial\alpha
\label{VI,1}
\eb

The covariant derivative:
\bb
D_\mu^\star = (\partial_\mu -iA_\mu)^\star
\label{VI,2}
\eb

The field strength tensor:
\bb
F_{\mu\nu} = [D_\mu^\star\ds D_\nu^\star ]=\partial_\mu A_\nu -\partial_\nu A_\mu -i[A_\mu\ds A_\nu ]
\label{VI,3}
\eb

The Lagrangian:
\bb
{\cal L} = -\frac 14 F^{\mu\nu}\star F_{\mu\nu}
\label{VI,4}
\eb

The field equations:
\bb
\partial^\mu F_{\mu\nu}  -i[A^\mu\ds F_{\mu\nu}] =0
\label{VI,5}
\eb

Consistency equations:
\bb
\partial^\nu [A^\mu \ds F_{\mu\nu}] = 0
\label{VI,6}
\eb

A schematic proof of the consistency condition:
\bea
&&[\partial^\nu A^\mu\ds  F_{\mu\nu}] +[A^\mu\ds \partial^\nu F_{\mu\nu}] = \label{VI,7} \\
&&=\frac i2 [ [A^\nu\ds A^\mu ]\ds F_{\mu\nu} ]+ i[A^\mu\ds [A^\nu\ds F_{\mu\nu}]]
\eea
We have used the field equations and the fact that $[F_{\mu\nu}\ds F^{\mu\nu}] = 0$. The terms left can now be
rearranged:
\bb
 [ [A^\nu\ds A^\mu ]\ds F_{\mu\nu} ] + [  [A^\mu \ds F_{\mu\nu} ]\ds A^\nu ] + [[F_{\mu\nu}\ds A^\nu]\ds A^\mu ]
\label{VI,8}
\eb
and vanish due to the Jacobi identity.

We found a conserved current:
\bb
j_\nu =[A^\mu\ds F_{\mu\nu}] , \qquad \partial_\nu j^\nu = 0 .
\label{VI,9}
\eb

\vskip1cm

{\bf 2) Electrodynamics with one charged spinor field.}

Tranformation law of the gauge field and the spinor field:
\bb
\delta\psi = i\alpha\psi,\quad \delta A_\mu =\partial_\mu\alpha
\label{VI,10}
\eb

Covariant derivative:
\bb
D_\mu^\star =(\partial_\mu -iA_\mu\star ),\quad D_\mu^\star \psi =(\partial_\mu -iA_\mu\star )\psi
\label{VI,11}
\eb

Field strength:
\bb
F_{\mu\nu} = \partial_\mu A_\nu -\partial_\nu A_\mu -i[A_\mu \ds A_\nu ]
\label{VI,12}
\eb

Lagrangian:
\bb
{\cal L} = -\frac 14 F^{\mu\nu}\star F_{\mu\nu} 
+ \bar\psi\star\gamma^\mu(i\partial _\mu + A_\mu\star )\psi - m\bar\psi \star\psi 
\label{VI,13}
\eb

Field equations:
\bea
&& \partial_\mu F^{\mu\rho} + i[A_\mu\ds F^{\rho\mu}] + \gamma^\rho \psi \star \bar\psi =0\nonumber \\[8pt]
&& (\gamma^\mu(\partial_\mu -iA_\mu \star ) + im)\psi = 0
\label{VI,14} \\[8pt]
&& \partial_\mu\bar\psi\gamma^\mu + i\bar\psi\gamma^\mu\star A_\nu - im\bar\psi =0
\nonumber
\eea

Consistency condition:
\bb
\partial_\rho ([A_\mu \ds F^{\rho\mu}]+\gamma^\rho \psi\star\bar\psi ) =0
\label{VI.15}
\eb
Proof: as before, the spinor terms have to be added
in the current and the field equations.

Current:
\bb
j^\rho =[A_\nu\ds F^\rho\nu] +\gamma^\rho \psi\star\bar\psi , \quad \partial_\nu j^\nu = 0 .
\label{VI,16}
\eb

\vskip1cm

\newpage

{\bf 3) Electrodynamics with several charged fields.}

We try to formulate a model with one vector potential
and differently charged matter fields as we do in the undeformed situation. This amounts to introduce an $U(1)$ gauge invariant action for the gauge potential and for the matter fields.

Let us consider the part of the vector potential first.

The transformation law is
\bb
\delta_\alpha A_\mu = \partial_\mu \alpha \label{VI,17} 
\eb

The covariant derivative
\bb
D_\mu^\star = (\partial_\mu -iA_\mu \star)
\label{VI,19} 
\eb
gives the following field strength tensor
\bb
F_{\mu\nu} =\partial_\mu A_\nu -\partial_\nu A_\mu -i[A_\mu\ds A_\nu ] .
\label{VI,19'}
\eb
As an invariant Lagrangian we choose
\bb
{\cal L}_{A} = -\frac 14 F^{\mu\nu}\star F_{\mu\nu} .\label{VI,20'}
\eb 

Next we consider the matter fields. They transform as follows
\bb
\delta_\alpha \psi^r = ig_r\alpha \psi^r .\label{VI,20''} 
\eb
The covariant derivative depends on the charge of the field it acts on:
\bb
D_\mu^\star \psi^r = (\partial_\mu -ig_rA_\mu\star )\psi^r .
\label{VI,18} 
\eb

The $U(1)$ gauge invariant action can be choosen as follows:
\bb
{\cal L}_\psi  =  \sum_r \bar\psi ^r\star\gamma^\mu(i\partial _\mu + g_r A_\mu\star )\psi^r - m_r\bar\psi^r \star\psi^r  .
\label{VI,19''}
\eb
As the total Lagrangian we take the sum
\bb
{\cal L} = {\cal L}_{A} + {\cal L}_{\psi} .\label{VI,21'}
\eb
It is $U(1)$ gauge invariant and it is a deformation of the usual electrodynamics with different charged fields. This Lagrangian now leads to the field equations:
\bea
&& \partial_\mu F^{\mu\rho} + i[A_\mu\ds F^{\rho\mu}] + \sum_r g_r\gamma^\rho \psi ^r\star \bar\psi^r =0, \nonumber \\[8pt]
&& (\gamma^\mu(\partial_\mu -ig_rA_\mu \star ) + im_r)\psi ^r= 0,
\label{VI,20} \\[8pt]
&& \partial_\mu\bar\psi^r\gamma^\mu + i\bar\psi^r\gamma^\mu\star g^r A_\nu  - im_r\bar\psi^r =0 .
\nonumber
\eea

The first of these equations gives rise to a consistency 
condition:
\bb
\partial_\rho (i[A_\nu \ds F^{\rho\nu}] + \sum_r g_r\gamma^\rho \psi^r\star\bar\psi^r ) =0 .
\label{VI.21}
\eb
From a direct calculation, using the field equations, follows:
\bea
&&\partial_\rho (i[A_\nu\ds F^{\rho\nu}]  + \sum_r g_r\gamma^\rho\psi^r\star\bar\psi^r )
\label{VI,22} \\
&&=-\sum_r( g_r^2-g_r)[A_\mu\ds \gamma^\mu\psi ^r\star\bar\psi ^r] .
\eea

The consistency condition is only satisfied if $g_r = g_r^2$ or $g_r = 1$. With one 
vector potential we can in a U(1) model only describe particles
with one  charge. There can be an arbitrary number of matter fields with this charge. 
This is different from the usual undeformed situation. There the comutator in (\ref{VI,19'}) vanishes 
and does not give rise to an inconsistency.

This is not surprising, we forgot that the vector potential
has at least to be envelopping algebra valued. This is 
demonstrated in the next example.

\vskip1cm

{\bf 4) Electrodynamics of a positive and a negative charged matter field.}

The gauge group is supposed to be $U(1)$ and the matter fields are in the multiplet that transforms as follows
\begin{equation}
\delta_\alpha\psi = i\alpha Q\psi, \quad\quad Q=
\pmatrix { 1 & 0\cr
0 & -1} .\label{VI33}
\end{equation}
As outlined in chapter 5, the gauge potential has to be in the same representation of the enveloping algebra as the matter fields are.

The enveloping algebra has two elements
\begin{equation}
I\> {\mbox{ and }}\> Q, \quad Q^2=1. \label{VI34}
\end{equation}
We generalize the transformation law (\ref{VI33}) to be enveloping algebra valued
\bb
\delta_\Lambda = i \Lambda\psi,\quad\quad \Lambda = \lambda_0(x)I + \lambda_1(x)Q.
\label{VI35}
\eb
The vector potential $A_\mu$ has the analogue decomposition
\bb
{\cal A}_\mu = A_\mu(x)I + B_\mu(x)Q. \label{VI36}
\eb
The covariant derivative is
\bb
D_\mu^\star \psi = (\partial_\mu -i{\cal A}_\mu\star )\psi 
= (\partial_\mu -iA_\mu(x)\star I -i B_\mu(x)\star Q)\psi 
\label{VI37}
\eb
The field strength can also be decomposed in the enveloping algebra
\bb
{\cal F}_{\mu\nu} = F_{\mu\nu}I + G_{\mu\nu}Q .\label{VI38}
\eb
From the definition of the field strength
\bb
{\cal F}_{\mu\nu} =\partial_\mu {\cal A}_\nu -\partial_\nu {\cal A}_\mu 
-i[{\cal A}_\mu\ds {\cal A}_\nu ] ,\label{VI39}
\eb
follows
\bea
&& F_{\mu\nu} =\partial_\mu A_\nu -\partial_\nu A_\mu -i[A_\mu\ds A_\nu ]
-i[B_\mu\ds B_\nu ], \nonumber\\
&& G_{\mu\nu} =\partial_\mu B_\nu -\partial_\nu B_\mu -i[A_\mu\ds B_\nu ]
-i[B_\mu\ds A_\nu ]. \label{VI40}
\eea

The matter fields couple to the vector potential via the covariant derivative
\bea
D_\mu^\star \psi &=& (\partial_\mu -i{\cal A}_\mu\star )\psi \nonumber\\
&=& (\partial_\mu -iA_\mu(x)\star I -i B_\mu(x)\star Q)\psi .
\label{VI41}
\eea
This leads to the Lagrangian
\bb
{\cal L} = -\frac 14 {\cal F}^{\mu\nu}\star {\cal F}_{\mu\nu} + \bar\psi\star\gamma^\mu(i\partial _\mu + {\cal A}_\mu\star )\psi - m\bar\psi \star\psi 
\label{VI42}
\eb
and the field equations
\bea
&&\frac{\delta{\cal L}}{\delta A_\rho}: \quad 
\partial_\mu F^{\mu\rho} +i[A_{\mu }\ds F^{\rho\mu}] + i[B_{\mu }\ds G^{\rho\mu}] 
+i\gamma^\rho\psi\star \bar\psi = 0, \nonumber\\
&&\frac{\delta{\cal L}}{\delta B_\rho}: \quad 
\partial_\mu G^{\mu\rho} +i[B_{\mu }\ds F^{\rho\mu}] + i[A_{\mu }\ds G^{\rho\mu}] 
+i\gamma^\rho\psi_A\star \bar\psi_B Q^{AB} = 0, \nonumber\\
&&\frac{\delta{\cal L}}{\delta\bar\psi}: \quad 
\gamma^\mu\big( \partial_\mu -i{\cal A}_\mu\star\big)\psi )+m\psi = 0 ,\nonumber\\
&&\frac{\delta{\cal L}}{\delta\psi}: \quad
\partial_\mu\bar\psi \gamma^\mu + i\bar\psi\gamma^\mu\star {\cal A}_\mu -m\bar\psi = 0 .
\label{VI43}
\eea
We obtain two consistency equations that render two transformation laws, in agreement with the extended symmetry (\ref{VI35})
\bb
j^\rho_{A} = i[A_\mu\ds F^{\rho\mu}] + i[B_\mu\ds G^{\rho\mu}]
+\gamma^\rho\psi_{A}\star\bar\psi_{A},
\label{VI44}
\eb
with
\bb
\partial_\rho j^\rho_{A} = 0 \label{VI45}
\eb
and
\bb
j^\rho_{B} = i[B_\mu\ds F^{\rho\mu}] + i[A_\mu\ds G^{\rho\mu}]
-i\gamma^\rho\psi_{A}\star\bar\psi_{B}Q^{AB}.
\label{VI46}
\eb

We learn that the deformed gauge theory leads to a theory with a larger symmetry structure, the enveloping algebra structure. This structure survives in the limit $\theta\to 0$. We find the corresponding conservation laws and gauge transformations needed for a consistent gauge theory. 

\vspace*{1.5cm}

{\Large{\bf  Acknowledgements}}

\vspace*{0.5cm}
I thank M. Buri\' c and M. Dimitrijevi\' c for very intesive discussions and for taking care of the manuscript. I also would like to thank the organizers of the Workshop who created a very stimulating atmosphere.

\end{document}